\documentclass[conference]{IEEEtran}
\IEEEoverridecommandlockouts
\usepackage{cite}
\usepackage{amsmath,amssymb,amsfonts}
\usepackage{algorithm}
\usepackage{algorithmic}
\usepackage{graphicx}
\usepackage{textcomp}
\usepackage{setspace}
\usepackage{float}

\def\BibTeX{{\rm B\kern-.05em{\sc i\kern-.025em b}\kern-.08em
    T\kern-.1667em\lower.7ex\hbox{E}\kern-.125emX}}
\begin{document}
\title{Joint Caching and Transmission in the Mobile Edge Network: A Multi-Agent Learning Approach}
\author{Qirui Mi, Ning Yang, Haifeng Zhang, Haijun Zhang,~\IEEEmembership{Senior Member,~IEEE}, Jun Wang}
\maketitle
\begin{abstract}

Joint caching and transmission optimization problem is challenging due to the deep coupling between decisions. This paper proposes an iterative distributed multi-agent learning approach to jointly optimize caching and transmission. The goal of this approach is to minimize the total transmission delay of all users. In this iterative approach, each iteration includes caching optimization and  transmission optimization. A multi-agent reinforcement learning (MARL)-based caching network is developed to cache popular tasks, such as answering which files to evict from the cache and which files to storage. Based on the cached files of the caching network, the transmission network  transmits cached files for users by single transmission (ST) or joint transmission (JT) with multi-agent Bayesian learning automaton (MABLA) method. And then users access the edge servers with the minimum transmission delay. The experimental results demonstrate the 
performance of the proposed multi-agent learning approach.
\end{abstract}


\section{introduction}
Internet of Thing (IoT) has become a communication paradigm, envisioning ubiquitous and seamless connectivity among users, data, and things \cite{IoT2020}. Globally, IoT connections are forecasted to grow 2.4-fold, from 2018 to 14.7 billion by 2023 \cite{WhitePaper}.

Successful deployment of large-scale edge IoT systems requires meeting the stringent quality-of-service (QoS) and delay requirements of massive users. Besides, the IoT networks are in general constrained by limited caching resources. An approach to tackling such an issue is to introduce the mobile edge caching (MEC) technique. In particular, the MEC extends storage resources to the network edge (e.g. base stations, WiFi access points, mobile devices). It can provide content or services needed by nearby users, reducing end-to-end latency and backhaul link traffic. In addition, due to the intensive deployment of wireless infrastructure, each user may be covered by multiple edge servers. Transmission strategies (i.e. User association strategy) affect the distribution of content requests to the edge servers, thus further affecting the efficiency of caching strategy. These motivate us to jointly consider caching and transmission problems.


Existing studies on learning-based caching strategies consisted of cooperative coded caching \cite{cache1}, caching replacement \cite{cache2}, and joint optimization caching and resource management\cite{ cache4}, such as joint caching and power allocation or joint caching and computing resource. In addition, the learning-based transmission strategies \cite{DQNAccess1, DQNAccess2} were widely studied to reduce latency by spectrum access, user association, power allocation. Users determined access decisions based on their current and past observations under the presence of spectrum sensing errors \cite{DQNAccess1}. The spectrum access optimal policy based on the deep Q-learning network (DQN) was studied to maximize the expected long-term number of successful transmission\cite{DQNAccess2}. However, existing researches mainly focus on caching or transmission, and the joint caching and transmission problem was not well studied. 



Some work on combining centralized caching policies and specific transmission methods\cite{centralized policy} for wireless communication network had been studied. However, it  is  challenging  to  directly  apply  them  to  joint  caching  and transmission optimization in a distributed manner. In general, caching strategy gets the scope of users to be served from transmission strategy, while transmission strategy is obtained based on cached files of multiple edge servers from caching strategy. Therefore, this non-stationary optimization problem motivates us to consider an iterative distributed multi-agent learning approach for joint caching and transmission optimization.

The contributions are summarized as follows: In this paper, we formulate a multi-agent learning approach to solve the joint caching and transmission problem in the edge IoT network. A multi-agent learning approach combines caching optimization and transmission optimization, aiming to minimize the total transmission delay. The MARL-based network with a multi-agent deep deterministic policy gradient (MADDPG) method is regarded as a caching network. It is developed to cache popular files, deciding which files to evict from the cache and which files to storage. Based on the cached files of the caching network, the transmission network performs hybrid transmission strategy with multi-agent Bayesian learning automaton (MABLA) method. The performance of the proposed multi-agent learning approach is compared with traditional cache strategies, transmission approaches and conventional reinforcement learning (RL) methods in terms of the transmission delay.   

The rest of this paper is organized as follows. Section II introduces the system model and formulates this problem. In Section III, we propose a multi-agent learning algorithm. Section IV analyzes proposed schemes. Section V performs numerical studies to evaluate the proposed schemes. Finally, Section VI concludes this paper.

\section{System model}
A MEC network is considered consisting of a cloud network $c$ and $E$ edge networks, denoted as a set $\mathcal{E} = \{ 1,2,...,E\}$. A set of users denoted by $\mathcal{U} = \{1,2,...,U\}$ request files each step according to Zipf popularity distribution, which is served by the cloud server or edge servers. The distribution of users is modeled as independent poisson point process (PPP) with density $\lambda $ in the MEC network \cite{PPP}. Each server is equipped with a cache to store files required by users. A file library is given by $\mathcal{F} = \{ 1,2,...,F\}$, and the size of each file is $s_f$ bits. Assume that the caches of the edge servers have equal capacity under the memory ${F_1}{s_f}$ bits, where $F_1$ is the number of files in each cache, ${F_1} < F$. The cache capacity is limited to
\begin{equation}
\begin{array}{l}
\sum\limits_{f = 1}^{F_1} {{x_{e,f}}} {s_f} \le {C_e},\forall f \in {\cal F}
\end{array}\label{eq:capacity restrict}
\end{equation}

The time horizon is divided into ${N^T}$ iterations, which are index by $T \in \{ 1,2,...,{N^T}\} $. Each iteration task has $N^t_1$ MARL steps and $N^t_2$ MABLA steps. Assume that the user requests a file at each step. The channel gains for the edge network and the cloud network are modeled as $h_{c,u}=g_{c,u}d_{c,u}^{(-{\alpha}/{2})}$ and $h_{e,u}=g_{e,u}d_{e,u}^{(-{\alpha}/{2})}$, respectively, where $g_{c,u}$ and $g_{e,u}$ are the Gaussian random variables, ${g_{c,u}} \sim C \mathcal{N}(0,1)$, ${g_{e,u}} \sim C\mathcal{N}(0,1)$.  The distances from cloud server $c$ to user $u$ and from edge server $e$ to user $u$ are $d_{c,u}$ and $d_{e,u}$.  The path loss factor is $\alpha$. 

\subsection{Cache Model}
Each edge network is equipped with a cache, it first checks whether requested file $f$ is cached in the edge network. If the requested file is available in the edge cache, then the edge server can transmit the file $f$ to the corresponding user without requesting them from the cloud server. To improve the hit ratio of the files, the caches of the edge servers need to be updated according to users’ preferences. Suppose that the file requested by a user will reach each step, which follows a Zipf popularity distribution with the skewness of $\upsilon$\cite{CacheModel2}. The caching indicator $\bf{x}$ is denoted as   
\begin{equation}
\begin{array}{l}
{{x_{e,f}} = \left\{ {\begin{array}{*{20}{c}}
1&{\text{cache}}\\
0&{\text{otherwise}}
\end{array}} \right.}
\end{array}
\end{equation}
where ${x_{e,f}}$ indicates if file $f$ is cached by edge server $e$.

\subsection{Communication Model}
In order to transmit the files requested by users to them as soon as possible, the cloud server and some edge servers will cooperate according to their caching situation. In the MEC network, the dense deployment of MEC infrastructures leads that some users have to be served by multiple MEC servers. These users are permitted to adopt hybrid transmission strategy to improve QoS. The success of transmission is based on the fact that requested files have been cached by edge servers. The user association indicator $\bf{y}$ is defined as  
\begin{equation}
\begin{array}{l}
{{y_{e,u}}= \left\{ {\begin{array}{*{20}{c}}
1&{\text{access}}\\
0&{\text{otherwise}}
\end{array}} \right.}
\end{array}
\end{equation}
user $u$ accesses the edge server $e$ for transmitting files $y_{e,u}=1$, otherwise, $y_{e,u}=0$. The requested indicator $\bf{z}$ is given by ${z_{u,f}} = \{ 0,1\} $, where $z_{u,f} = 1$ means user $u$ requests the file $f$, otherwise, $z_{u,f} = 0$. Assume that the edge server $e$ covers $U^e$ users, which share a channel and interfere with each other. The channel gains are sorted as $|{h_{e,1}}| \ge ...  \ge |{h_{e,{U^e}}}|$ for edge $e$. The interference is reduced with successive interference cancellation (SIC) technology \cite{ImperfectSIC}. 

For the ST, user $u$ covered by edge server $e$ will first send the file request to the edge server $e$. If requested file $f$ is available, and user $u$ will be served by edge server $e$. The achievable downlink  data rate from edge server $e$ to user $u$ is given by
\begin{equation}
\begin{array}{l}
{{R_{e,u}}\! \! = \! \! \sum\limits_{f = 1}^{{F}} {{B_{e,u}}lo{g_2}(1\!  +\!  \frac{{{x_{e,f}}{y_{e,u}}{z_{u,f}}|{h_{e,u}}{p_{e,u}}{|^2}}}{{\sum\limits_{i = u + 1}^{U^e} {{x_{e,f}}{y_{e,i}}{z_{i,f}}|{h_{e,i}}{p_{e,i}}{|^2}}  + \sigma _u^2}})} }
\end{array}\label{eq:ST data rate}
\end{equation}
where $B_{e,u}$ is the channel bandwidth for edge server $e$ to user $u$. The user association variable ${y_{e,i}}$ represents whether user $i$ associates with edge server $e$. The requested indicator ${z_{i,f}}$ shows if user $i$ requests file $f$. The power of user $u$ and user $i$ from edge network $e$ are ${p_{e,u}}$ and ${p_{e,i}}$, respectively. In general, some users are served by multi-servers
since the ultra-dense deployment of the edge network. Suppose that there are $E^u$ edge servers to transmit the cached files to user $u$. For JT model, the downlink data rate for user $u$ is defined as
\begin{equation}
\begin{array}{l}
{{R_u} \! \!= \! \!\sum\limits_{e = 1}^{{E}} {\sum\limits_{f = 1}^{F} {{B_{e,u}}lo{g_2}(1\! \! +\! \! \frac{{{x_{e,f}}{y_{e,u}}{z_{u,f}}|{h_{e,u}}{p_{e,u}}{|^2}}}{{\sum\limits_{i = u + 1}^{{U^e}} {{x_{e,f}}\!{y_{e,i}}\! \!{z_{i,f}}|{h_{e,i}}{p_{e,i}}{|^2}}\!   +  \!\sigma _u^2}})} } }
\end{array}\label{eq:JT data rate}
\end{equation} 
The equations \eqref{eq:ST data rate} and \eqref{eq:JT data rate} represent the transmission data rate of either ST mode or JT mode for user $u$ in the edge networks.
If the related edge servers cannot provide requested file $f$ by user $u$, user $u$ will send request to the cloud server $c$. An indicator function $\mathbb{I}(\sum\limits_{e = 1}^E {{x_{e,f}}{y_{e,u}}{z_{u,f}}}  \ge 1)$ is denoted to reflect whether requested file $f$ is transmitted successfully by an edge server. The downlink transmission data rate from cloud server $c$ to user $u$ is stated as
\begin{equation}
\begin{array}{l}
{{R_{c,u}} \! \!= \! \!{B_{c,u}}lo{g_2}(1 \! \!+\! \! \frac{{(1 - \mathbb{I}(\sum\limits_{e = 1}^E {{x_{e,f}}{y_{e,u}}{z_{u,f}}}  \ge 1))|{h_{c,u}}{p_{c,u}}{|^2}}}{{\sum\limits_{i = 1,i \ne u}^{U^c} {(1 - \mathbb{I}(\sum\limits_{e = 1}^E {{x_{e,f}}{y_{e,i}}{z_{i,f}}}  \ge 1))|{h_{c,i}}{p_{c,i}}{|^2}}  + \sigma _u^2}})}
\end{array}\label{eq:cloud data rate}
\end{equation} 
where ${p_{c,u}}$ and ${p_{c,i}}$ are transmit powers from cloud server $c$ to user $u$ and user $i$. The channel gains from cloud server $c$ to user $u$ and user $i$ are ${h_{c,u}}$ and ${h_{c,i}}$. The transmission bandwidth from cloud $c$ to user $u$ is $B_{c,u}$. The cloud server $c$ transmit signal to ${U^c}$ users, $1 \le U^c \le U$.  
\subsection{Delay Model}
In the subsection, the transmission delay for the cloud network and edge network is calculated. The transmission delay for ST mode and JT mode by edge servers is given by
\begin{equation}
\begin{array}{l}
{{D^{E2U}} = \sum\limits_{u = 1}^U {\sum\limits_{e = 1}^E {\sum\limits_{f = 1}^{{F}} {\frac{{{x_{e,f}}{y_{e,u}}{z_{u,f}}{s_f}}}{{{R_{e,u}}}}} } } }
\end{array}\label{eq:edge data rate}
\end{equation} 

The transmission delay for all users from the cloud network is 
\begin{equation}
\begin{array}{l}
{{D^{C2U}} = \sum\limits_{u = 1}^U {\sum\limits_{f = 1}^F {\frac{{(1 - \mathbb{I}(\sum\limits_{e = 1}^E {{x_{e,f}}{y_{e,u}}{z_{u,f}}}  \ge 1)){s_f}}}{{{R_{c,u}}}}} } }
\end{array}\label{eq:cloud delay}
\end{equation} 
where $s_f$ is the size of the file $f$. 

\section{multi-agent Caching and Transmission}
In this section, a joint caching and transmission problem is studied. Unlike existing researches, we utilize an iterative multi-agent method to minimize the total transmission delay. In each iteration, the caching network learns the caching policy according to users' preference to optimize caching variables $\bf{x}$ in the first step. Furthermore, to minimize the transmission delay, the transmission network optimizes access variables $\bf{y}$ by selecting optimal transmission mode in the second step. The transmission delay minimization problem can be formulated as

\begin{equation}
\begin{array}{l}
{\mathop {\min }\limits_{\{ {\bf{x}},{\bf{y}}\} } \;{D^{E2U}} + {D^{C2U}}}
\end{array}\label{eq:Model function}
\end{equation}
\begin{equation}
\begin{array}{l}
\begin{aligned}
& \text{s.t.}
& & {C1:{\bf{x}} \buildrel \Delta \over = {\{ {x_{e,f}}\} _{e,f}},{x_{e,f}} \in \{ 0,1\} ,\forall e,\forall f},\\
&&& {C2:{\bf{y}} \buildrel \Delta \over = {\{ {y_{e,u}}\} _{e,u}},{y_{e,u}} \in \{ 0,1\} ,\forall e,\forall u},\\
&&& {C3:\sum\limits_{f = 1}^F {{x_{e,f}}{s_f}}  \le {C_e},\forall e,\forall f},\\
&&& {C4:\sum\limits_{e = 1}^E {\sum\limits_{u = 1}^{U^e} {{p_{e,u}} + \sum\limits_{u = 1}^{U^c} {{p_{c,u}} \le P} } } }.
\end{aligned}
\end{array}\label{eq:Model constraints}
\end{equation}
where constraints (C1)-(C2) limit the caching variables and the user association variables to binary variables. Constraint (C3) describes the cache of each edge server has a limited capacity $C_e$. Constraint (C4) ensures that the power consumption is limited to the peak power of the system. The problem is challenging to solve for the following reasons:
\begin{itemize}
\item Conventional solutions require complete parameters for this problem, which may be unaffordable to obtain full parameters in a large-scale IoT system.
\item The objective \eqref{eq:Model function} involves the caching variables $\bf{x}$ and user association variables $\bf{y}$, which is non-convex.

\item The feasible set of constraints \eqref{eq:Model constraints} is non-convex as a result of binary variables $\bf{x}$ and $\bf{y}$.

\item The problem in \eqref{eq:Model function} is combinatorial, which is difficult to solve optimally. In particular, a brute-force method requires the computational complexity ${\mathcal O}({(F)^E} + {(E)^U})$, so it's impractical to obtain the optimal joint caching and transmission strategy with large-scale users and files.
\end{itemize}

The aforementioned challenges motivate us to use a multi-agent learning approach to separately optimize the large-scale parameters of the caching network and the transmission network. A multi-agent learning structure is shown in 
Fig. 1.

\begin{figure}
    \small
    \centering
    \includegraphics*[width=90mm]{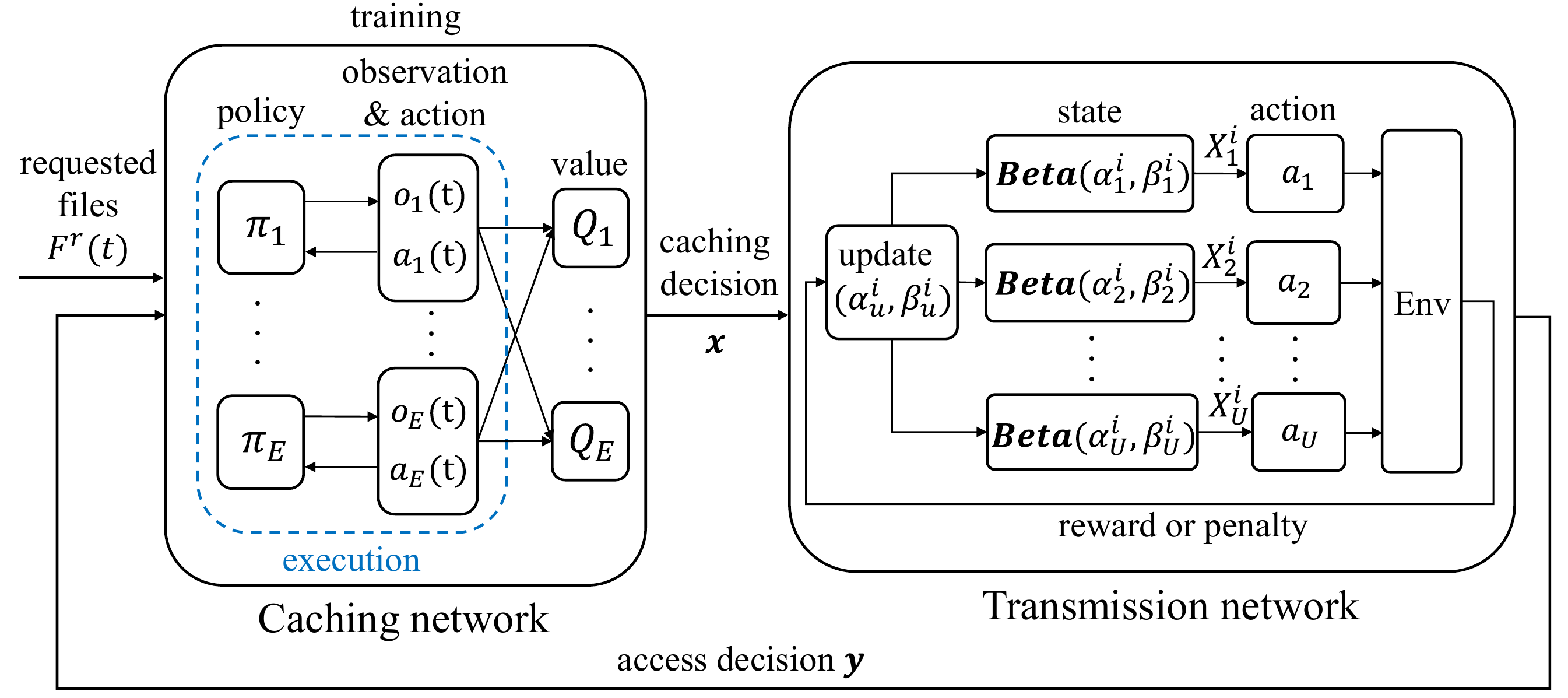}
    \caption{The multi-agent learning structure}
    \label{fig:1}
\end{figure}


The multi-agent learning structure includes two parts: the caching network and the transmission network. The left one is the caching network which predicts and caches the files interested by users based on MARL. The input of the network is the requested files and user access information at last timestep and the output is caching decision $\bf{x}$. Each edge server is regarded as an agent to learn a caching strategy ${\pi _e},\forall e$. The right one is the transmission network. Based on the caching decision ${\bf{x}}$, the transmission network learns a transmission strategy by MABLA method for users served by multiple edge servers. The users make access decision ${\bf{y}}$ based on the state information that is represented by Beta distribution, whose parameters are updated utilizing the feedback (reward or penalty) of the environment.

\subsection{MARL-based Edge Caching}
In order to solve the optimization problem \eqref{eq:Model function}, a joint caching and transmission strategy has been proposed to optimize the caching variables $\bf{x}$ and user access variables $\bf{y}$ for the given $\bf{z}$. In the caching network, multiple edge servers are considered and each of them is regarded as an agent. The edge server interacts with its environment at every step. Each agent cannot obtain the full environment state and they make cache decisions in a distributed manner. Therefore, the caching problem can be modeled as a Markov game\cite{multi-agent}, which is defined with a tuple $\langle{\mathcal{S},\mathcal{A}, \mathcal{P}, \mathcal{R}, \gamma}\rangle$ for this multi-agent case. $\mathcal{S}$ denotes the state space, $\mathcal{A}$ denotes the joint action space, $\mathcal{P}:\mathcal{S}\times \mathcal{A} \to PD(\mathcal{S})$ denotes the transition function, $\mathcal{R}:\mathcal{S}\times \mathcal{A} \to \mathcal{R}$ denotes the reward function, and $\gamma$ denotes the discount factor.

{\bf{State $\mathcal S$}}: The state of the environment contains the requested files by users $F^r(t)$ and the cached files by edge servers $ F^c(t)$ at step $t$, which is denoted by
\begin{equation}
\begin{array}{l}
\mathcal{S}=\{ F^r(t), F^c(t)\}
\end{array}
\end{equation}
\begin{equation}
    \begin{array}{l}
F^r(t) = \{F^r_1(t),...,F^r_E(t)\}
\end{array}
\end{equation}
\begin{equation}
    \begin{array}{l}
F^c(t) = \{F^c_1(t),...,F^c_E(t)\}
\end{array}
\end{equation}
where $F_e^r(t) \in {\mathbb{R}^{U^e}}, F_e^c(t) \in \mathbb{R}^{F_1},\forall e$ is the vector of $U^e$-dimension and $F_1$-dimension, respectively. The requested files by the covered users of edge server $e$ is ${F_e^r(t) = \{f_e^1(t),....,f_e^{{U^e}}(t)\}}$, where $f_e^u(t)$ is the requested file of user $u$ served by edge server $e$, ${f_e^u}(t) \in \{ 1,2,...,F\} ,\forall e$. The cached files by edge sever $e$ is ${F_e^c(t) = \{f_e^1(t),....,f_e^{{F_1}}(t)\}}$, where $f_e^{F_1}(t)$ is the cached file served by edge server $e$, $f_e^{F_1}(t) \in \{ 1,2,...,F\}$.

{\bf{Observation $\mathcal O$}}: Assume that each agent is unable to obtain the full environment state information. The edge server $e$ only can observe its requested files of users in its own covered area that is given by 
\begin{equation}
\begin{array}{l}
\begin{array}{*{20}{l}}
{{\cal O} = \{ {o_e}(t)|{o_e}(t) = \{ F_e^r(t)\} \} }
\end{array} 
\end{array}
\end{equation}

{\bf{Action $\mathcal A$}}: Suppose that caches of agents are full and cached files are indexed by $\{ 1,...,{F_1}\}$. Each agent has to select a deleted file from $F_1$ cached files and an added file from $F$ library files when updating its cache. The action is represented as
\begin{equation}
\begin{array}{l}
\begin{array}{*{20}{l}}
{{\cal A} = \{ a_1(t),...,{a_E}(t)|{a_e}(t) \in \{ 0,1,...,{F_1}F\} \} }
\end{array}
\end{array}
\end{equation}

Therefore, we can get the update information as
\begin{equation}
\begin{array}{l}
\begin{array}{l}
f_e^c(t) = \left\{ {\begin{array}{*{20}{c}}
0 ,\qquad &{{a_e}(t) = 0}\\
1 ,\qquad &{\text{otherwise}}
\end{array}} \right.\\
f_e^d(t) = {a_e}(t)/F \qquad
f_e^a(t) = {a_e}(t)\% F
\end{array}
\end{array}
\end{equation}
where $f_e^c(t)$ shows whether to update the cached files of edge server $e$. If the files are updated $f_e^c(t)= 1$, otherwise $f_e^c(t)= 0$. The index of the deleted file from its cache is $f_e^d(t)$ and the index of the added file from files library $\mathcal F$ is $f_e^a(t)$.

{\bf{Reward $\mathcal R$}}: The goal of the agent is to find a caching strategy $\mathcal \pi$ to minimize the transmission delay according to evicted and cached files. The reward function is given by
\begin{equation}
\begin{array}{l}
\mathcal{R} = \{r_1(t),...,r_E(t)\}
\end{array}
\end{equation}
\begin{equation}
\begin{array}{l}
{{r_e}(t) = 1/(\sum\limits_{u = 1}^U {\sum\limits_{f = 1}^{{F}} {\frac{{{x_{e,f}}(t){y_{e,u}}(t){z_{u,f}}(t){s_f}}}{{{R_{e,u}}(t)}}} })} 
\end{array}
\end{equation}

{\bf{Transition $\mathcal P$}}:
The transition shows the rules of taking action $a(t)$ from $s(t)$ to ${s(t+1)}$. For example, if the cached files in edge server $e$ are $\{1,2,3\}$ at step $t$, the action of agent $e$ is to delete file $1$ and add file $4$, then the cached files at step $t+1$ will change to $\{4,2,3\}$.

{\bf{Discount factor $\gamma$}}:
The cached files at the next state ${s(t+1)}$ is affected by the action $a(t)$, so $\gamma$ is introduced to the reward function to discount the future reward. 

In the caching network, the MADDPG method\cite{maddpg} is introduced to learn caching policy, which is based on a actor-critic model. In the actor network, the policies of the multi-agent are defined as $\pi  = \{ {\pi _{{\theta _1}}},...,{\pi _{{\theta _E}}}\}$ with parameters $\theta  = \{ {\theta _1},...,{\theta _E}\} $, the gradient of the expected reward for edge server $e$ is 
\begin{equation}
\begin{array}{l}
\begin{array}{*{20}{l}}
{{\nabla _{{\theta _e}}}J\left( {{\theta _e}} \right) = {E_{{a_e}\sim{\pi _e}}}\left[ {{\nabla _{{\theta _e}}}log{\pi _{\theta_e}}\left( {{a_e}|{o_e}} \right)Q_e^\pi \left( {s,a;{\theta _e}} \right)} \right]}
\end{array} \label{eq:GradientActor} 
\end{array}
\end{equation}
where ${Q_e^\pi \left( {s,a;{\theta _e}} \right)}$ is the action-value function, which is updated by loss function of the edge server $e$. 
\begin{equation}
\begin{array}{l}
\begin{array}{*{20}{l}}
{{\cal L}\left( {{\theta _e}} \right) = E\left[ {{{\left( {Q_e^\pi \left( {s,a;{\theta _e}} \right) - {y_e}} \right)}^2}} \right]}
\end{array}\label{eq:LossCritic}  
\end{array}
\end{equation}
where ${y_e}$ is the approximate target action-value function for the critic network with parameters $\phi  = \{ {\phi _1},...,{\phi _E}\} $. The target action-value function is given by
\begin{equation}
\begin{array}{l}
\begin{array}{*{20}{l}}
{\begin{array}{*{20}{l}}
y_e = r_e + \gamma Q^{\pi'}_e (s',a';\phi_e)|_{a'_e=\pi_{\phi_e}(o_e)}
\end{array}}
\end{array}
\end{array}
\end{equation}
where $\pi'=\lbrace{\pi_{\phi_1},...,\pi_{\phi_E}}\rbrace$ is a set of the target strategy.  
\subsection{MABLA-Based Transmission}
Based on the caching decision of the caching network, the transmission network makes the user association decision. The $U^E$ users covered by multi-edge servers need to select a transmission mode, such as ST or JT. The JT means that the requested file is transmitted by two or more edge servers. The ST means the cached file is transmitted by only one server. 

The MABLA method is proposed in order to apply Bayesian learning automaton(BLA) method\cite{BLA} to multiple users. According to \textbf{Theorem 1}, the MABLA method can converge to an optimal value with two actions case and it has low computing complexity than other learning automatons. Motivated by above advantages, a MABLA-based transmission network is proposed to optimize the user association variables. Transmission decisions have two arms, i.e. ST and JT. The core of MABLA is the beta distribution that generating Bayesian estimates of the reward probability of each action. In the MABLA-based transmission network, the ST is selected by user $u$ as Arm 0, while the JT is denoted as Arm 1. The probability density function of the beta distribution is represented by
\begin{equation}
\begin{array}{l}
f(x;\alpha,\beta)=\frac{x^{\alpha-1}(1-x)^{\beta-1}}{\int_{0}^{1}{u^{\alpha-1}(1-u)^{\beta-1}du}},x\in[0,1]
\end{array}
\end{equation}
where $\alpha $ and $\beta $ are the parameters of the beta distribution. The state of user $u$ at step $t$ is given by 
\begin{equation}
\begin{array}{l}
{s_u}(t) = (\alpha _u^0(t),\beta _u^0(t),\alpha _u^1(t),\beta _u^1(t))
\end{array}
\end{equation}

The parameters of first beta distribution and second beta distribution are $(\alpha _u^0,\beta _u^0)$ and $(\alpha _u^1,\beta _u^1)$. The action of user $u$ is denoted as 
\begin{equation}
\begin{array}{l}
{{a_u}(t) = \left\{ {\begin{array}{*{20}{c}}
{Arm 0}&{{X^0_u}(t) > {X^1_u}(t),}\\
{Arm 1}&{{\text{otherwise.}}}
\end{array}} \right.}
\end{array}
\end{equation}
where $X^i_u(t)$ is sampled from \textbf{Beta}$(\alpha^i_u,\beta^i_u)$ of user $u$. After taking the action, the parameters of the Beta distribution are updated as follows:
\begin{equation}
\begin{array}{l}
{\left\{ {\begin{array}{*{20}{c}}
{\alpha _u^i(t + 1) = \alpha _u^i(t) + 1}&{\text{ reward},}\\
{\beta _u^i(t + 1) = \beta _u^i(t) + 1}&{\text{ penalty}.}
\end{array}} \right.}
\end{array}\label{eq:Beta distribution parameters}
\end{equation}
where $i = 0,1$. If user $u$ with the selected arm has a lower delay than another arm, user $u$ obtains a reward, otherwise gets a penalty.\\
\textbf{Theorem 1}: When $t\to\infty$, MABLA is able to converge to only choosing the optimal hybrid transmission strategy $\pi^*$, i.e. $\lim_{t\to\infty}p_{\pi^*}\to1$.\\
\emph{Proof}:
According to \cite{theorem1}, the feedback of Arm $i$ for user $u$ provides a reward with probability $\delta^i_u$, the expected value $\mathbb{E}(X^i_u)=\frac{\alpha^i_u(t)}{\alpha^i_u(t)+\beta^i_u(t)}$ approaches $\delta^i_u$ over time, i.e.:
\begin{align}
    \mathbb{E}(X^i_u) &=\frac{\alpha^i_u(t)}{\alpha^i_u(t)+\beta^i_u(t)}=\delta^i_u\\
    \beta^i_u &=\frac{(1-\delta^i_u)\alpha^i_u(t)}{\delta^i_u}
\end{align}
The probability of choosing its optimal arm $i$ for user $u$ is
\begin{align}
    p^i_u &= \frac{(\beta^i_u(t))!(\alpha^{1-i}_u(t))!}{(\beta^i_u(t)+\alpha^{1-i}_u(t))!}
    = \frac{(\frac{1-\delta^i_u}{\delta^i_u}\alpha^i_u(t))!(\alpha^{1-i}_u(t))!}{(\frac{1-\delta^i_u}{\delta^i_u}\alpha^i_u(t)+\alpha^{1-i}_u(t))!}
\end{align}
\\if $t\to\infty$,
\begin{align}
  \delta^i_u \to 1,\frac{(1-\delta^i_u)}{\delta^i_u}\alpha^i_u\to 0  \\
  p^i_u \to \frac{0!(\alpha^{1-i}_u(t))!}{(0+\alpha^{1-i}_u(t))!} \to1
\end{align}
Thus when $t\to\infty$, the probability of converging to the optimal hybrid transmission strategy $\pi^*$ is given by\\
\begin{equation}
\begin{array}{l}
p_{\pi^*}=\prod \limits_{u=1}^U{p^i_u} \to1
\end{array}
\end{equation}
The proof is completed.

\section{Analyzes the Proposed multi-agent Approach} 
A description of the proposed multi-agent learning approach for joint caching and transmission is shown in Algorithm 1.

\begin{algorithm}[H]
\caption{A multi-agent Learning Approach}
\begin{algorithmic}[1]
\STATE  Initialize state $\mathcal S$ for the caching network.
\FOR     {${T} = 1$ to ${N^T}$}
\STATE \textbf{First step: MARL-based Edge Caching}
\FOR     {${t} = 1$ to ${N^t_1}$}
\STATE For each edge server $e$, select caching action $a_e(t)$ with observation $o_e(t)$ by evaluate network.
\STATE Get reward $r_e(t)$, next state $s_e(t+1)$ for each agent.
\STATE Store transitions for edge servers in replay buffer $\mathcal{D}$.
\STATE Set $s_e(t) = s_e(t + 1)$ and update cache variables $\textbf{x}$.
\FOR {edge server $e=1$ to $E$}
\STATE Sample a random minibatch from $\mathcal{D}$. 
\STATE Update evaluate network parameters by \eqref{eq:GradientActor}\eqref{eq:LossCritic}.
\ENDFOR
\STATE Update target network parameters for each edge server.
\ENDFOR
\STATE \textbf{Second step: MABLA-based hybrid transmission}
\STATE  Initialize $s_u(t)$: $\alpha^i_u:=\beta^i_u:=1$.
    \FOR{${t} = 1$ to $N^t_2$}
        \FOR{${u} = 1$ to ${U^E}$}
            \STATE Generate two values $X_0$ and $X_1$ randomly from Beta distribution \textbf{Beta}$(\alpha^0_u,\beta^0_u)$ and \textbf{Beta}$(\alpha^1_u,\beta^1_u)$. 
            \STATE If $X_0>X_1$, choose ST, otherwise JT.
            \STATE Compute feedback value.
            \STATE Update $s_u(t)$ by \eqref{eq:Beta distribution parameters}.
            \ENDFOR
        \ENDFOR  
        \STATE Update $\mathcal S$, $\bf{y}$.
\ENDFOR      
\end{algorithmic}
\end{algorithm}

\begin{itemize}
\item\textit{Caching network}: 
For caching network, the state of the caching network is initialized including the requested files of users and cached files in edge servers firstly. Each edge server selects caching action $a_e(t)$ from caching network. Then, the server will obtain next state $s_e(t+1)$ and reward $r_e(t)$. These transitions for all edge servers are stored into replay buffer $\mathcal{D}$ as experience data. Each agent updates its caching network parameters in turn. The parameters of target network will be updated in each step.
\item\textit{Transmission network}: 
For the transmission network, user $u$ samples two values from $\textbf{Beta}(\alpha^i_u, \beta^i_u)$ and selects the action with the greater value. Then user $u$ gets a reward or penalty and updates the parameters of $\textbf{Beta}(\alpha^i_u, \beta^i_u)$ according to \eqref{eq:Beta distribution parameters}. With the number of iteration increases, it converges to the optimal transmission strategy.
\end{itemize}
\section{Simulation}
\begin{figure}[t]
    \small
    \centering
    \includegraphics*[width=76mm]{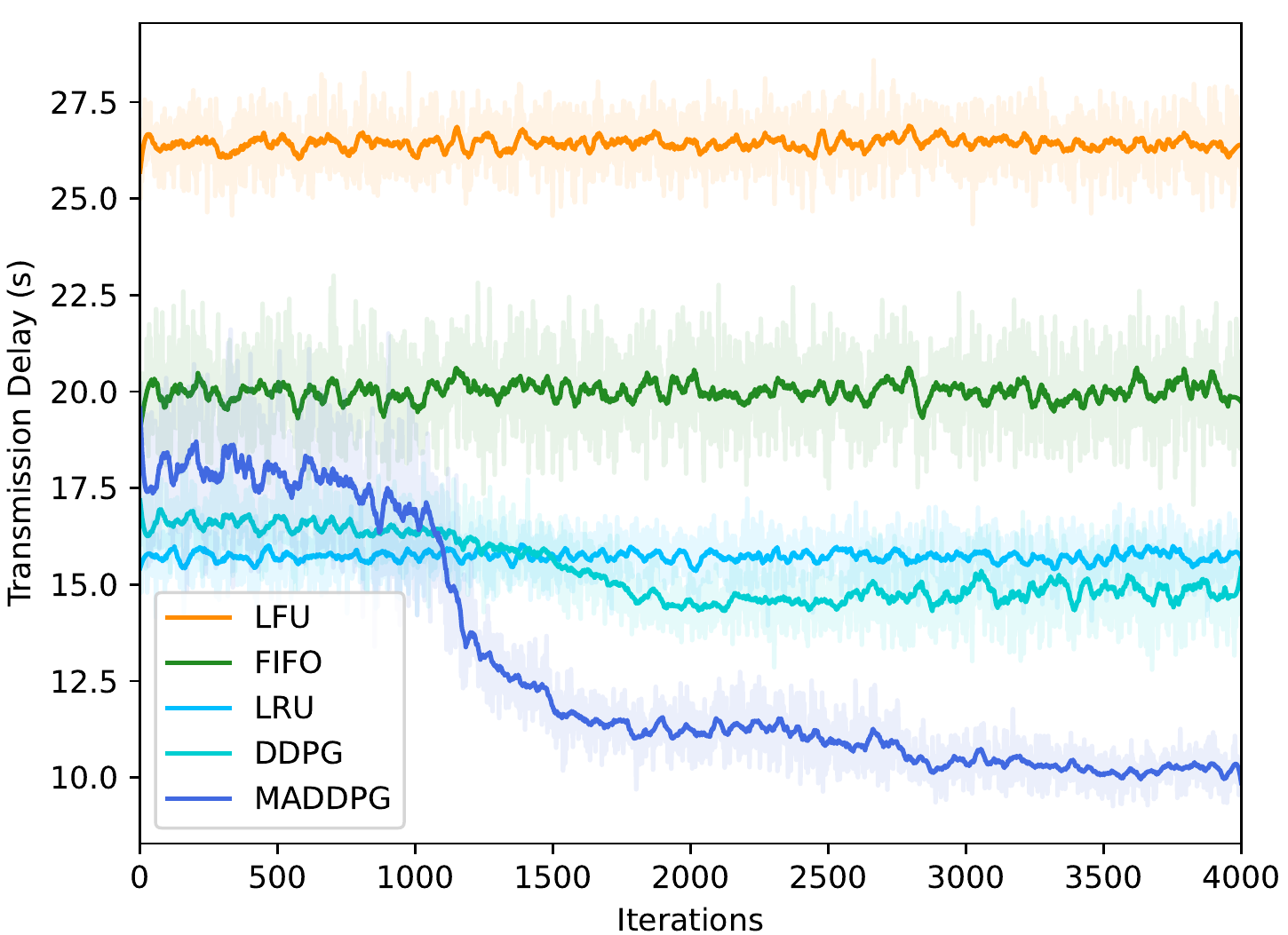}
    \caption{The performance of MARL-based caching on transmission delay}
    \label{rl-cache}
\end{figure}

In this section, the performance of the proposed multi-agent learning approach for the mobile edge network is verified. In the simulations, we consider three intersecting circular cells of three edge servers with the same radius $r = 100$m, i.e., $E=3$. Each edge server has a $10$MB cache capacity and the size of each file is $1$ MB. Supposed that 20 users are sampled by independent PPP with the density $\lambda=200 /km^2$. The skewness of Zipf popularity distribution $\upsilon=1.2$, the peak power of the system $P=39.953$W, the bandwidth $B_{e,u}=B_{c,u}=4.5$MHz, $\forall e, \forall u$, files' number $F=50$, the distance $d_{c,u}=3Km, \forall u$, path loss factor $\alpha=4$. MARL hyperparameters are shown in Table I.
\begin{table}[H]
\caption{MARL Hyperparameters of System.}
\begin{center}
\begin{spacing}{1.07}
\begin{tabular}{l|l}
\hline\hline
{\bf Hyperparameter} & {\bf Value}\\
\hline
Multi-agent iterations ${T}$ & $4000$ \\
\hline
MARL steps $N^t_1$ & $75$\\
\hline
MABLA steps $N^t_2$ & $50$\\
\hline
Replay memory & $10^5$ \\
\hline
Learning rate & $1.5\times10^{-4}$ \\
\hline
Decay rate & $0.001$ \\
\hline
Discount rate & $0.95$ \\
\hline
Initial exploration  & $0.03$ \\
\hline
Final exploration & $0.0$ \\
\hline
Batch size & $512$ \\
\hline
Hidden dimension & $128$\\
\hline \hline
\end{tabular}
\end{spacing}
\end{center}\label{table:1}
\end{table}
Fig.\ref{rl-cache} shows the transmission delay versus the number of iterations for the five caching algorithms. Assume that the transmission network adopts the JT mode for the five algorithms. The superiority of the proposed MARL-based caching algorithm with JT is demonstrated by comparing with three traditional caching strategies with JT and a single-agent reinforcement learning(SARL)-based caching algorithm with JT. From Fig. 2, it is seen that the least recently used (LRU), least frequently used (LFU), and first in first out (FIFO) caching strategies with JT don’t have a downward trend as the number of iterations increases since they are all fixed strategies. Only SARL-based and MARL-based caching algorithms can reduce the total transmission delay with the increasing iterations. However, compared with SARL-based caching, MARL-based caching has a larger delay reduction and gets the best performance. In the beginning, the proposed MARL-based caching algorithm with JT has a higher transmission delay than LRU caching with JT and SARL-based caching with JT. After 1500 iterations, it gets the least delay among all caching algorithms. The performance of proposed method is optimized according to the dynamic requested files for the users.
\begin{figure}[t]
        \small
        \centering
        \includegraphics*[width=76mm]{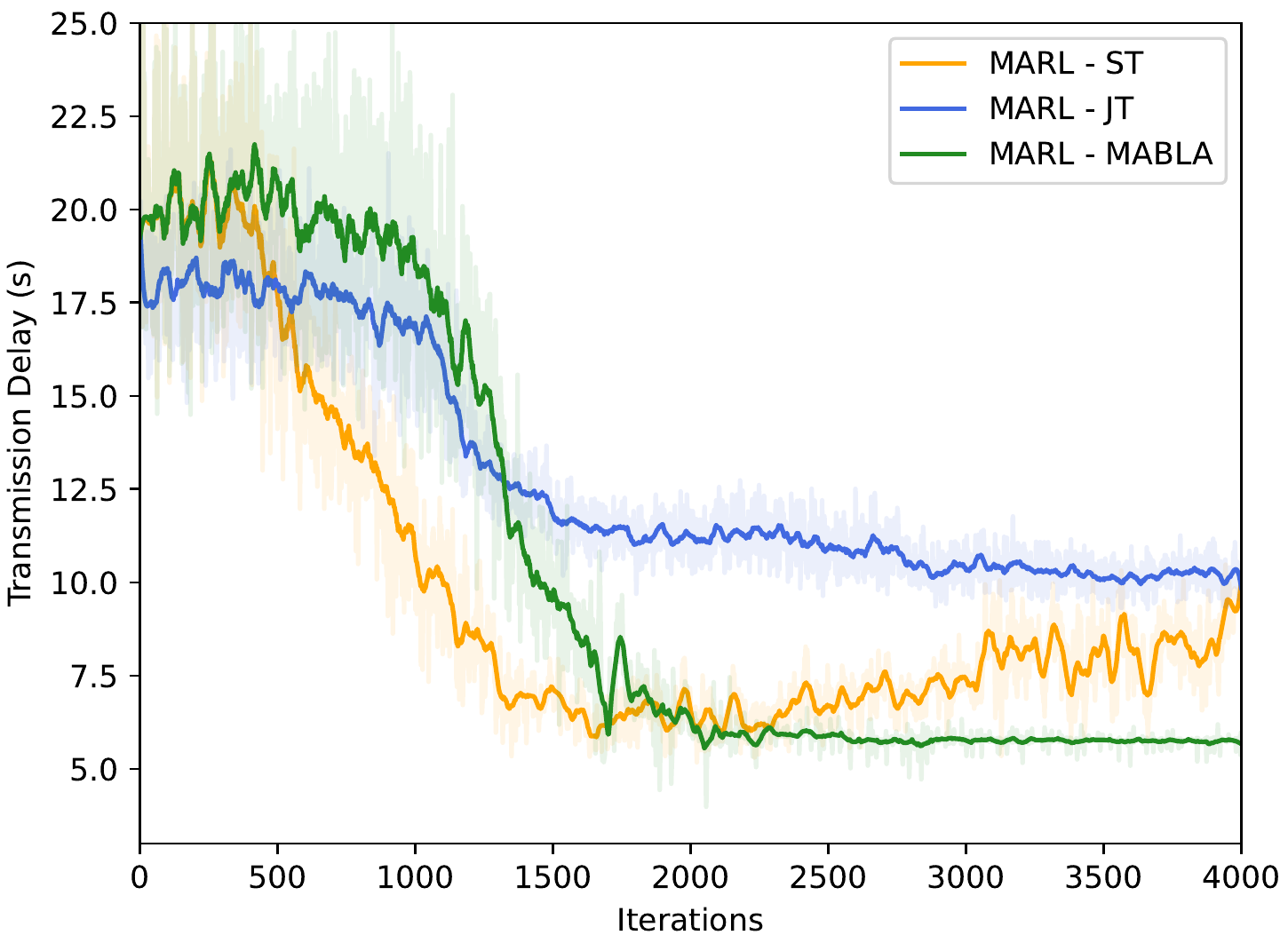}
        \caption{The performance of MARL-MABLA on transmission delay}
        \label{fig:2}
\end{figure}

Fig.\ref{fig:2} shows the transmission delay versus the number of iterations for the three transmission strategies. Supposed that the caching network applies MARL-based caching strategy for MARL-JT, MARL-ST, and MARL-MABLA algorithms. For the transmission network, the MARL-JT algorithm and MARL-ST algorithm adopt ST and JT for all users, respectively. Before 1300 iterations, the MARL-MABLA algorithm shows the highest transmission delay than the others. Compared with MARL-ST and MARL-JT algorithms, the MARL-MABLA algorithm converges to the lowest delay after 2000 iterations and keeps stable.


\section{Conclusion}
In this paper, an iterative multi-agent learning approach is proposed to minimize the total transmission delay of all users by optimizing caching and transmission in the mobile edge network. In each iteration, a caching network is optimized considering multi-edge servers with MADDPG method firstly. Based on the caching files, a transmission network is proposed considering multi-users with MABLA method to transmit the cached files by hybrid transmission strategy. Simulation results show the proposed multi-agent learning approach achieves the best performance among the existing caching approaches (i.e., FIFO, LFU, and LRU), transmission approaches (i.e., ST and JT), and the conventional RL method(i.e., SARL) in terms of transmission delay. However, with the number of file types and edge servers increases,  the stability and convergence of the proposed algorithm are affected in the complex non-stationary environment. It is worth studying the large-scale joint caching and transmission. 

\section*{Acknowledgements}
Ning Yang and Haifeng Zhang are partly supported by the Strategic Priority Research Program of Chinese Academy of Sciences, Grant No. XDA27030401.

\end{document}